# 3D Vector Piezoresponse Imaging with Interferometric Atomic Force Microscopy

Roger Proksch[1] and Ryan Wagner[2]

[1]Asylum Research an Oxford Instruments Company, Santa Barbara, CA, 93117 USA

[2]School of Mechanical Engineering, Purdue University, West Lafayette, IN, 47907 USA

ABSTRACT: Forces acting between an Atomic Force Microscope (AFM) tip and sample are three dimensional. Despite this, most AFM force measurements are confined to one or two dimensions. Extending AFM force measurements into three dimensions has previously required complex, difficult and time-consuming workflows. Here, we demonstrate an accurate, interferometric method for quantifying the full, three-dimensional response of an AFM tip to localized forces. We demonstrate this approach on a series of piezoelectric materials and show that this approach yields quantitative 3D measurement independent of the sample orientation beneath the tip. This approach simplifies existing, angle-resolved piezoresponse force microscopy (PFM) techniques. Our measurements benefit from the greatly reduced noise floor ($\approx 5\text{fm}/\sqrt{\text{Hz}}$) and intrinsic accuracy of our interferometric measurements. One important result is that the vertical piezo sensitivity ($d_{\text{eff},z}$, units of pm/V) was systematically 2-3x larger than the in-plane piezo sensitivities ($d_{\text{eff,lat}}$). A simple analysis of vertical and lateral contact stiffnesses, due to the difference in the Young (vertical) and Shear (lateral) sample moduli $d_{\text{theory},z}/d_{\text{theory,lat}} \approx 2.5$, in good agreement with the measurements. While this work was confined to ferroelectric materials, it provides a general workflow and framework for other AFM based mechanical measurements.

1. Introduction

The nanoscale mechanical properties of materials determines their performance in variety of critical applications including energy storage[1,2] and production[3] 2D materials,[4,5] beyond Moore's law computing materials[6] including neuromorhphic computing[7] and low-power electronics.[8] In particular, shear strain and modulus offer avenues for enhanced control over material properties at the nanoscale. As device dimensions shrink and traditional scaling reaches its limits, shear strain engineering enables fine-tuning of electronic properties,[9] altering band structures, and introducing anisotropy in material response.[10] This control is crucial for devices such as field-effect transistors, spintronic systems, and quantum devices, where optimized carrier mobility, reduced energy dissipation, and tailored quantum states are needed to overcome limitations in performance and efficiency. Moreover, the local shear modulus plays a critical role in the mechanical stability of nanoscale devices that operate under high strain environments, enhancing durability and flexibility, and allowing devices to maintain functionality under strain.

Stress and strain are two of the most important concepts in materials science and are intrinsically three-dimensional (3D). Macroscopic 3D measurements of material properties requires careful treatement and special approaches; a relatively undeveloped area for nanoscale properties. In the following, we focus on using the atomic force microscope (AFM),[11] to quantify the 3D functional response of ferroelectric materials. This formalism can also be extended to accurately map the 3D mechanical properties of materials and the 3D properties of the tip-sample contact.

The AFM, because its combination of high resolution, environmental versatility and myriad combined measurement modalities has become a mainstay in nano-science and technology. Typically, AFMs are based around a flexible micromachined cantilever with a sharp tip at the end that localizes the interactions with a sample. Most AFM measurements only measure one displacement component of the 3D tip motion vector, the vertical displacement component, $\delta z$, of the cantilever away from its equilibrium position. The force can then be estimated from the deflection observable through Hooke's law, $F_z = k_z \delta_z$, where $k_z$ is the spring constant of the cantilever at the tip location and $\delta_z$ is the displacement of the tip. It is also common to measure in-plane displacements of the cantilever tip using the lateral twisting of the cantilever[12] using the optical beam deflection (OBD) technique,[13] discussed more completely below. The OBD technique allows simultaneous measurements of both the vertical and lateral response of the AFM tip. Notably, despite over thirty years since the invention of the OBD, reliable and reproducible calibration of the detector remains a very active field of research. To paraphrase Munz,[14] while numerous calibration approaches exist, further refinement is needed to enhance accuracy, hopefully to metrological standards, and be easy and accurate enough to be broadly acceptable in the AFM community.

Angle-resolved piezoresponse force microscopy (AR-PFM) [15], [16], [17], [18], [19], [20], [21], [22], [23], [24] was developed to provide high-resolution, multidirectional mapping of polarization domains in ferroelectric and piezoelectric materials. This is particularly vital in ferroelectrics and multiferroics, where domain configurations can significantly affect material properties and functions and where different phases or orientations coexist. AR-PFM has several challenges that can impact its effectiveness in characterizing ferroelectric materials. Experimental and instrumental challenges include tip wear and degradation, where repeated scanning at multiple angles can cause tip deterioration, affecting resolution, piezoresponse sensitivity, and measurement repeatability. [17] Additionally, cantilever artifacts such as torsion and buckling can distort the extracted domain orientation maps.[20], [23], [24] Achieving precise sample rotation is another concern, as misalignment in manual or automated rotation can lead to mismatches between successive images, necessitating careful image registration. [18]. Furthermore, electromechanical crosstalk between vertical and lateral piezoresponse components can introduce artifacts, particularly in materials with strong anisotropic behavior.[19] Beyond instrumentation, data processing and interpretation also present significant hurdles. Complex image registration is required to align multiple images taken at different angles, often necessitating advanced image-processing techniques such as polynomial corrections or machine learning-based registration.[18] The technique also faces material and sample constraints that can alter domain behavior. Thin film and substrate influences, such as strain relaxation, charge screening and charge injection can modify the observed ferroelectric domain configurations. This can be especially in materials with high leakage currents, such as $BiFeO_3$ that may exhibit distorted piezoresponse signals due to charge accumulation at domain walls. Finally, there are practical considerations for applying AR-PFM in research and industry. The technique is time-intensive, as acquiring multiple PFM images at different angles significantly increases scanning time, limiting throughput in large-scale studies. Perhaps most significantly, the overwhelming majority of the AR-PFM report the in-plane responses in terms of "arbitrary units", a reflection of the difficulty of calibrating the full, three-dimensional sensitivity OBD measurements of the cantilever motion. (eight out of the ten references at the beginning of this paragraph reported exclusively in "volts" or arbitray units). Despite these challenges, AR-PFM remains a valuable tool for probing complex ferroelectric domain structures that can benefit from an accurate and easy to implement calibration technique. A recent exception to this uncalibrated approach is in the work of Alikin et al. for lateral[25] and longitudinal[26] in-plane measurements. In their work, Alikin et al. pointed out that both the in-plane and vertical piezoresponse can cause an apparent vertical response since they both cause the cantilever to bend. They showed that by varying the OBD spot position, the relative contributions of the in-plane and vertical could be controlled. This controlled mixing allowed the vertical and in-plane contributions to be quantitateively separated. In the following, we build on a new spot-position dependent interferometric approach to PFM[27, 28] that allows faster, easier and more accurate measurement of the full, three dimensional functional response of ferroelectric materials and sets the stage for future measurements of the 3D response of many other functional materials.

The first Atomic Force Microscope (AFM) used a tunneling detector, followed by various interferometric detection methods. In 1986, Martin and Wickramasinghe[29] introduced an AFM with a heterodyne laser interferometer. In 1987, the OBD solution was invented,[13] providing a relatively inexpensive, low-noise detection method that has since become nearly ubiquitous in the AFM field, with most commercial AFMs based around it. In this method, laser light is focused onto the back of the cantilever, reflected off and detected by a position-sensitive detector. The output signal from the PSD responds when the cantilever undergoes angular displacement, directly proportional to the deflection of the cantilever.

Despite the dominance of the OBD method, there continued to be improvements in interferometric detection. While the noise floor of these interferometers attained impressive levels below 10fm/rtHz [30],[31] they suffered from very limited measurement ranges, in the range of a few nm. In 2015, we demonstrated that an interferometric detection system (IDS) could directly probe the motion of an AFM cantilever tip based around a laser vibrometer that had a much higher range of motion.[27] This system has enabled separation of electrostatic interactions and quantification of the inverse piezoresponse in various materials. [32],[33],[34], [35],[36],[37] Very recently, we have released a quadrature phase differential interferometer (QPDI) that has a very low noise floor ($\approx 5\ fm/\sqrt{Hz}$) while maintaining a very large dynamic range (>1um).[38]

Early in the history of PFM, researchers took advantage of lateral detection of OBD based AFMs to image both the vertical and in-plane lateral piezoresponse.[39] By rotating the sample under the probe, a full, three dimensional vector Vector Piezoresponse Force Microscopy[40] (Vector PFM) image can be estimated. Vector PFM measurements based on OBD measurements require precise calibration and data treatment to ensure that the independent x, y and z PFM components. In addition, re-constructing three-dimensional PFM images of polarization is strongly affected by frictional sliding of the tip and longitudinal bending (sometimes inexactly referred to as buckling) of the cantilever.[40, 41] In existing approaches, this process requires careful rotation and relocation of the sample to the same measurement area,[16] which can be difficult and time-consuming. Additionally, thermal drift and repositioning inaccuracies can cause slight differences in the tip position, complicating the alignment of corresponding data points. [20],[19],[18],[16],[17] Finally, those existing works typically struggle to yield quantitative estimation of the in-plane piezoresonse amplitudes, image usually appear with "arbitrary units", underscoring the challenges of calibrating the OBD sensitivities.

In this work, we demonstrate how interferometric AFM can be extended to quantitatively measure the full, three dimensional response of AFM probes interacting with a sample by mapped the full 3D Piezoresponse Force Microscopy (PFM) response on a variety on ferroelectric samples. PFM measures the local electromechanical response by applying a modulated potential to the cantilever tip and detecting the mechanical response at the same frequency, primarily due to the converse piezoelectric effect.

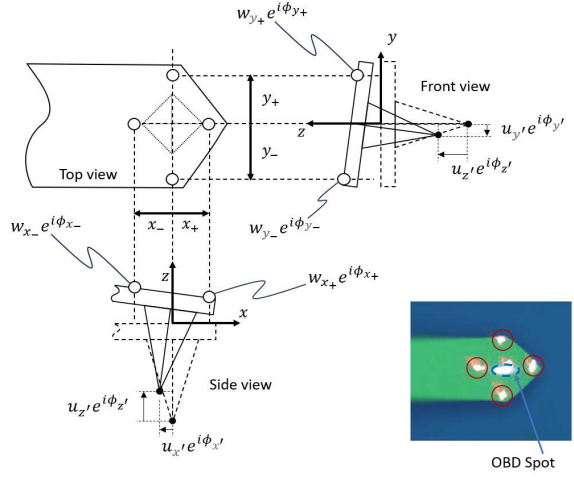

Figure 1 a) shows the end of a cantilever with a coordinate system (x, y, z) positioned above a sample (x', y', z'). The displacement of the cantilever on its top surface is described by three . The cantilever can also bend around the longitudinal axis (lateral) The inset shows a composite image with the OBD spot in the center and the composite interferometer spots at the four cardinal positions discussed in the supplemental material.

## 2. Interferometric Vector PFM

Single frequency motion of the tip in contact with the sample surface is described by a vector

$$\boldsymbol{u} = \left(\widehat{x'}u_{x'}e^{i\phi_{x'}} + \widehat{y'}u_{y'}e^{i\phi_{y'}} + \widehat{z'}u_{z'}e^{i\phi_{z'}}\right)e^{i\omega t}. \qquad 1$$

Similarly, we can describe motion of the cantilever along the interferometer measurement axis (z) in response to the tip motion, as

$$W(x,y,t) = w(x,y)e^{i(\omega t + \phi(x,y))} \qquad 2$$

Although we can make measurements at many points on the back of the cantilever, for this work, we define displacement measurements at four "cardinal points" as $w_{x_+} \equiv w(x_+, 0), w_{x_-} \equiv w(x_-, 0), w_{y_+} \equiv w(0, y_+)$ and $w_{y_+} \equiv w(0, y_-)$ We also make the measurement points equidistant around the tip location, $|x_+| = |x_-| = |y_+| = |y_-|$. If we assume the cantilever is a rigid body, in-plane motion of the tip is translated into tilting of the cantilever body. Then, Equations (1) and (2) can be solved for the amplitudes and phases of the different components (full derivation in the supplemental material) given by

$$u_{x'}e^{i\phi_{x'}} = \frac{(r_0 L - x_-)w_{x_+}e^{i(\phi_{x_+} - \phi_{inst})} - (r_0 L + x_+)w_{x_-}e^{i(\phi_{x_-} - \phi_{inst})}}{2 r_0 L G}, \qquad 3$$

$$u_y e^{i\phi_{y'}} = \frac{w_{y_+}e^{i\phi_{y_+} - \phi_{inst}} - w_{y_-}e^{(i\phi_{y_-} - \phi_{inst})}}{2G}, \qquad 4$$

$$u_{z'}e^{i\phi_{z'}} = \frac{w_{x_+}e^{i(\phi_{x_+}-\phi_{inst})} + w_{x_-}e^{i(\phi_{x_-}-\phi_{inst})}}{2}, and \quad 5$$

$$u_{z'}e^{i\phi_{z'}} = \frac{w_{y_+}e^{i(\phi_{y_+}-\phi_{inst})} + w_{y_-}e^{i(\phi_{y_-}-\phi_{inst})}}{2}. \quad 6$$

In these expressions, $L$ is the length of the cantilever (base to tip) and $r_0 = 2/3$ is the dimensionless ratio that determines the effective lever arm length for the zeroth mode of the cantilever. We have made the measurement location equi-distant from the tip location and introduced a geometric factor $G = |x_+|/h = |x_-|/h = |y_+|/h = |y_-|/h$, where $h$ is the height of the tip. In addition, the interferometer spot positions were chosen such that $G \approx 1$. The instrumental, frequency-dependent phase offset $\phi_{inst}$ is assumed to be constant for all of the components and can be determined a number of ways as discussed in the literature. [42, 43, 44] The remainder of the parameters are experimental observables that vary pixel to pixel. If the sample is rotated 90 degrees, as is shown in Figures 2b), 2d), 3b) and 3d) below, transformed equations are given in the supplemental materials (S19-S21).

Equations 3-6 (or S19-S21) provide a full 3D description of the tip motion at the modulation frequency in terms of experimental observables, Note that the z' component can be estimated from the weighted average of the x+, x- or y+, y- pairs for both the un-rotated (Equations 3-6) and rotated (Equations S19-S21) measurements. Post-acquisition comparisons of these independent estimates allows testing of the time-stability of the sample response and measurements. Supplemental Figure S4 shows an example where a large AC bias voltage, close to the coercive field causes changes in the vector components of the sample domain structure.

In practical measurements with real cantilevers, there is variation in the tip location and dimensions on the cantilever body. These can be characterized with optical or SEM images of the probes. In many cases, these dimensional uncertainties are dominant but even if they are neglected rarely amount to more than 10% uncertainty in the displacement measurements (see supplemental materials). While this is significant from many types of physical measurements, one should keep in mind that many current OBD-based calibrations may have uncertainties that are vastly larger; >100% error is not uncommon, especially in electromechanical measurements where the presence of uncontrolled in-plane frictional forces and long-range body-electrostatic forces can lead to signals that often dwarf the signal contributions from the true vertical electromechanical displacements. [45, 23, 34, 28] It is also interesting to note that while the vertical displacements (Equations 5 and 6) depend on plan-view uncertainties of both the tip and measurement spot locations, since the in-plane measurements relay on differences of the interferometric measurements $w_j$ and $\phi_j$, where $j = x+, x-, y+, y-$, they are largely independent on errors in the plan view location of the tip.

In the experimental measurements shown below, we tested the variability associated with the choice of the relative orientation of the sample with the cantilever. Specifically, we rotated samples by 90 degrees so that, for example, the x' component of the sample was both measured with the detection spots separated along the longitudinal axis of the cantilever (Equation 3) and, in a separate measurement after it was rotated with the spots separated along the lateral axis of the cantilever (Equation 4).

As is common, we define an effective converse piezoresponse sensitivity as $d_{eff,j} = (|u_j|/V_{ac}) \cdot sin(\phi_j - \phi_{inst})$, where the amplitude is $|u_j| = \left(\sqrt{Re(u_j e^{i(\phi_j - \phi_{inst})})^2 + Im(u_j e^{i(\phi_j - \phi_{inst})})^2}\right)$ and the phase as $\phi_j = tan^{-1}(Im(u_j e^{i\phi_j})/Re(u_j e^{i\phi_j})) + \phi_{inst}$, where $j = x', y'$ or $z'$ and $V_{ac}$ is the bias modulation voltage amplitude. We chose $\phi_{inst}$ to rotate the measured phasors along either the real or imaginary axes of the complex plane. This allowed separation of positive and negative polarization orientations in histograms (see for

example the phasor representation in S). The final result of the calculations 3-6 or 7-10 are plotted as images of the piezoresponse, $d_{eff,j}$. The absolute orientation was calibrated by locally poling the sample and measuring the phase shift over the poled region.

We used the following experimental workflow to map the vector components:

1. Choose a region of the sample and perform four separate interferometric images with the beam positioned at the cardinal points $(x_+, 0), (x_-, 0), (0, y_+)$ and $(0, y_-)$ as shown in Figure 1.
   a. Make the fast-scanning axis for the imaging along the axis perpendicular to the spot position axis (e.g., scan the sample parallel to the y-axis of the cantilever when making the $w_{x_-}$ and $w_{x_-}$ measurements etc…)
   b. Adjust scan speeds if necessary to ensure that trace and retrace response curves match to within acceptable experimental uncertainty.
2. Process the measurements according to Equations 3-6 if $x' \parallel x$ (sample y-axis parallel to cantilever axis) or S19-S21 if $x' \parallel y$ (sample perpendicular to cantilever axis).
3. As noted in the literature,[46],[44] in single frequency PFM, there is typically a frequency-dependent phase shift associated with the control electronics and signal processing of the modulated electromechanical response. This is equivalent to a rotation of the In the case of single frequency PFM, that means we can estimate a phase shift that orients the collective phasor of a series of measurements to be aligned with the in-phase or quadrature axis. This is illustrated in Supplemental Figure S2.

## 3. Experimental Results

The output of this workflow is shown in Figures 2 and 3 for BFO and PZT respectively. The imaging steps (1 and 3 above) in workflow were executed automatically with a MacroBuilder™ routine. MacroBuilder™ is a measurement automation interface in the Asylum Research AFM software (see https://www.youtube.com/watch?v=en8McD40gN8 for an overview of this feature). The sample rotation and recentering was accomplished using optical fiducials followed by fine tuning of the scanning locations using the x' and y'-offsets in the AFM. An advantage of the above workflow compared to previous AR-PFM approaches is that the number of images required is greatly reduced, improving tip wear and other time dependent degradation of the response.

One of our initial questions was whether or not there was a systematic difference in measurements made along the long axis (longitudinal, x-axis) of the cantilever and measurements made in the transverse (lateral, y-axis) axis of the cantilever. To test for this rotational variance we rotated the samples of Figures 2 and 3 by 90° around the center of the image area. We performed steps 1-3 that the rotated reference frame. Explicitly, in the rotated reference frame, measurements made at cardinal points $(x_+, 0)$, $(x_-, 0)$ are sensitive to the sample $y' -$ and $z' -$components of the sample strain, while $(0, y_+), (0, y_-)$ are sensitive to the sample $x' -$ and $z' -$components of the sample strain. In addition to being sensitive to differences in-plane stiffnesses of the cantilever tip, [26] comparisons of the original and rotated vector components tests our "rigid body model" of the end of the cantilever.

Figure 2 hows the results of our workflow applied to an epitaxial [100] BFO sample.[47] As noted above, a) and c) were made on an unrotated sample and b) and d) were made with the sample rotated by 90 degrees (see the sample edge on the bottom of a) and c) and on the right side of images b) and d). We performed steps 1-3 of the above workflow in both of the rotatation states, producing, in total, four independent estimates of the vertical (z') piezoresponse and two each of the in-plane (x' and y') components. In the case of the BFO, inspection of all three spatial component shows excellent correlation, independent of the sample rotation angle. The z' images in Figure 2 show some texture, most likely due to some small uncertainties in the location of the tip. Inspection of the image histograms for z' (plotted to the right of the images, each with the same scale as shown in the color bars) show good agreement. The small variations in amplitude of z' may also be due to small errors in the plan position of the AFM tip (denoted with the diamond shape in Figure 1) or small errors in the positioning of the spots during the measuring step.

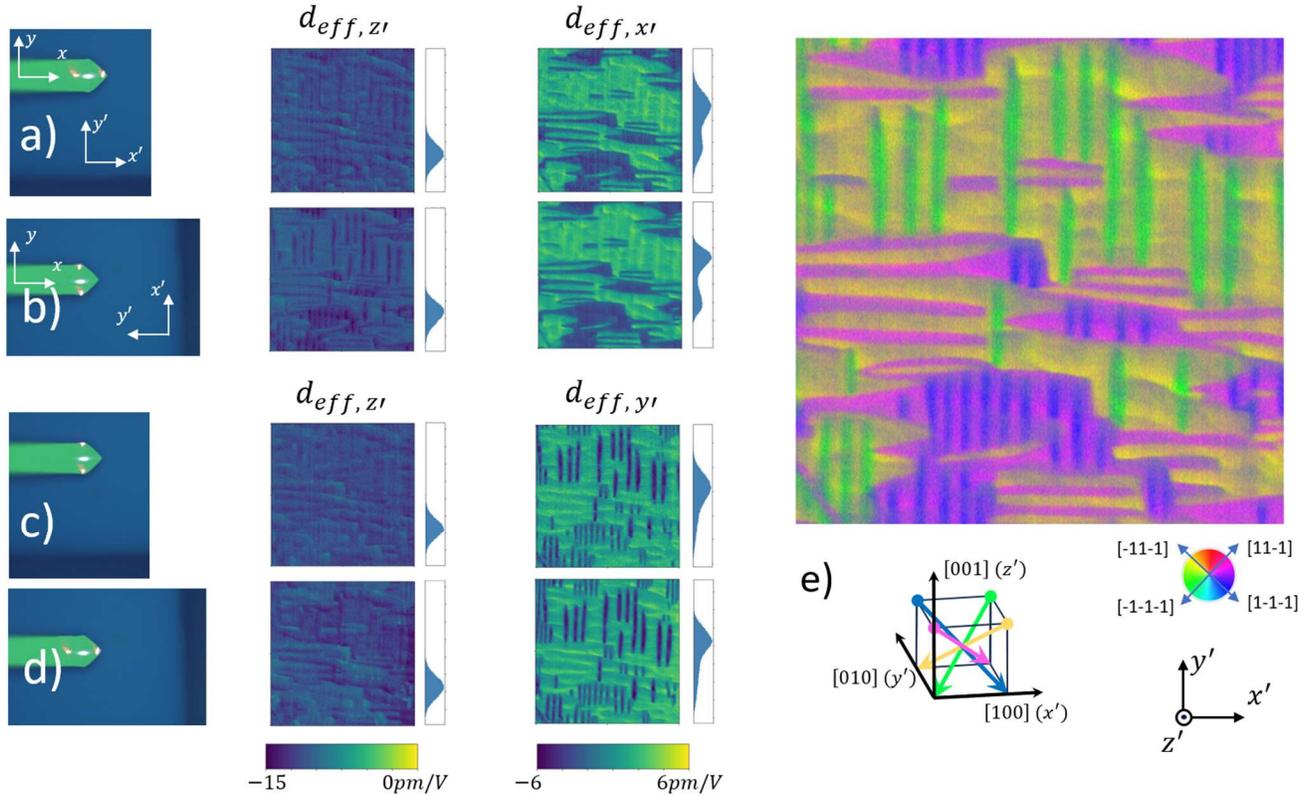

Figure 2. a)-d) show estimated vector piezoresponse ($d_{eff,\ j}$, where $j = x', y'$ or $z'$) component images of BFO made at two different sample orientations using the workflow described in the text. Specfically, a pair of interferometric measurements were made for each case a) through d), allowing the estimation of a pair of vector components, one in-plane (longitudinal for a) and d), lateral for b) and c)) and two vertical. 2 a) and b) show two independent measurements of x' and z' pairs, while c) and d) show two independent measurements of y' and z' pairs. Histograms of the amplitude components are plotted adjacent to the images, the ranges are the same as the color bars at the bottom. e) illustrates the polarization components along the crystollagraphic axes of the sample and the color-coded in-plane vector components derived from the measurement data in a) and c). These data were acquired with a Cypher IDS with separated OBD and interferometric measurement spots. The scan range was 7.5µm.

In Figure 2, the polarization was found to be in one of four states along the body diagonal of the unit cell. In all cases, the z-component was oriented into the plane, while there were four in-plane components as shown in Figure 2e). This is in contrast to earlier OBD-based AR-PFM work, where researchers found multiple polarization variants in an unswitched sample. [48], [19]

Figure 2e) also shows a scaled RGB image where each pixel is colored by the sum of the normalized magnitude of the in-plane amplitudes with clearly differentiated domains. Note that from the histogram of values in Figure 2a-d), the span of the vertical sensitivities averaged $\approx -12.5 \pm 2 pm/V$, while the in-plane is $\approx \pm 5 \pm 1 pm/V$. The in-plane responses all had both x' and y' components while the deff, z' components were uniformly negative. This enabled us to color-code the in-plane ($d_{eff,\ x'}$ and $d_{eff,\ y'}$) map with four distinct polarization vectors in along the unit cell body diagonals. The results shown in Figure 2, specifically, the agreement between components measured at 0 and 90 degree sample rotations are consistent with our treatment of the end of the cantilever as a rigid body. Conveniently, it also implies that full 3D vector measurements can be made with reasonable confidence without rotating the sample.

Note that all of the images $d_{eff,\ x'}$, $d_{eff,\ y'}$, and $d_{eff,\ z'}$ in Figure 2 show small modulations consistent with crosstalk between different components. We hypothesize that this may originate from small errors in the plan-view tip location as discussed in the supplemental materials. This may be particularly evident when comparing, for example $d_{eff,\ z'}$ in Figure 2b) to $d_{eff,\ y'}$ in Figure 2c), where the vertical component $d_{eff,\ z'}$ appears to have clear features similar to the in-plane $d_{eff,\ y'}$ image. This implies that there is a small longitudinal ($\delta x_{off}$) error in the Figure 2b) lateral measurements that then results in a small contribution from the longitudinal motion of the tip. Note that, despite this crosstalk being visible in the image, it results in uncertainty that is still within the error budget in Table S1.

Figure 3 is a series of sequential images of polished PZT (details in the supplemental material) at two different rotational orientations with respect to the cantilever axes – 0 and 90 degrees. The component images were calculated using equations (3)-(6). Similarly to the BFO results shown in Figure 2, the $d_{eff,\ z'}$ components of the PZT response shown in Figure 3a)-d) are strikingly similar, as are the in-plane components $d_{eff,\ x'}$ and $d_{eff,\ y'}$. The measurements at 0 degrees, 3a) and 3c) were made first, followed by 3b) and d). The estimated in-plane measurements ($d_{eff,\ x'}$ and $d_{eff,\ y'}$) are also very uniform in Figure 3, showing very weak dependence on the sample orientation. As with the data shown in Figure 2, this validates our rigid tip model described in Figure 1 above.

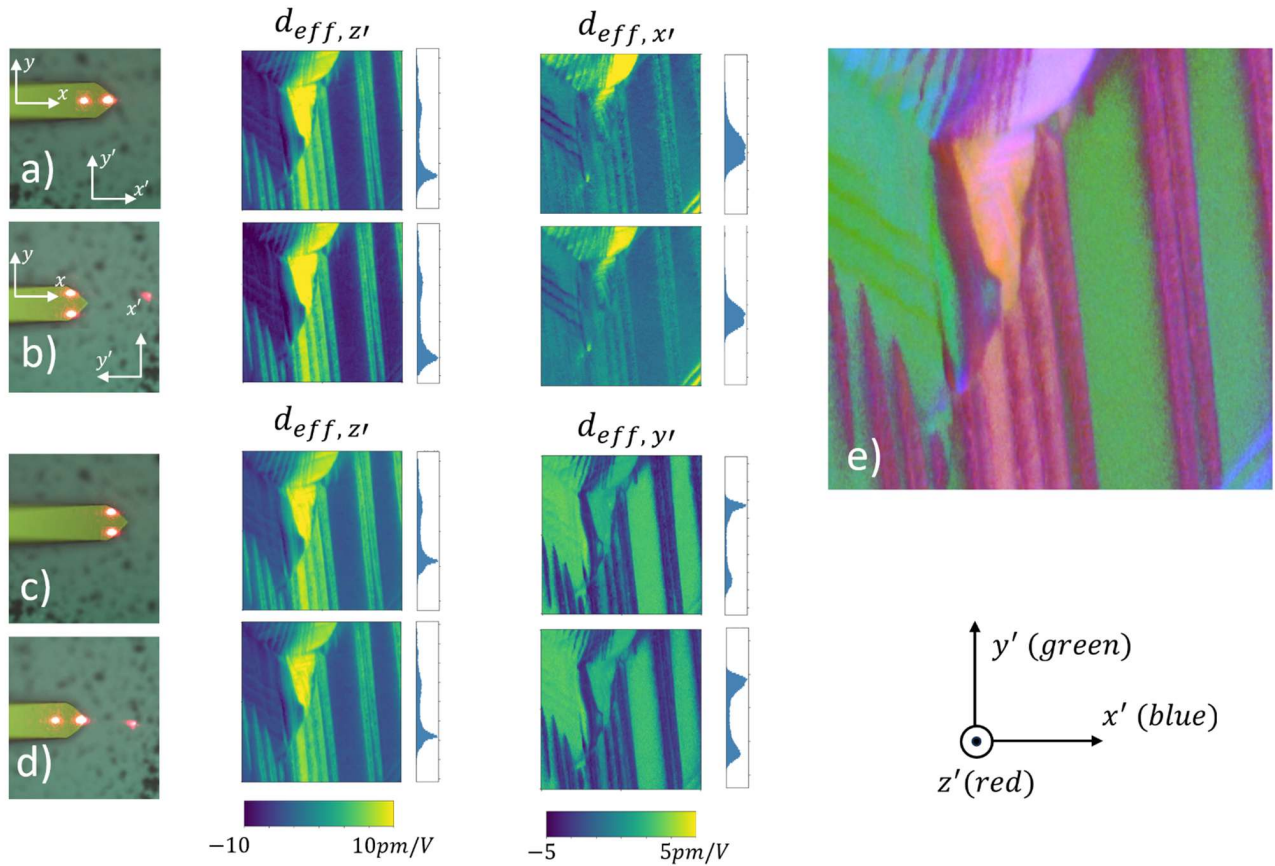

Figure 3. Sequential vector piezoresponse images of a PZT sample at two different sample orientations. The composite optical images in a) - d) show the interferometer spot locations on the top of the cantilever along with the orientation of the sample below it. The measurements were made following the workflow described for Figure 2. For reference, the optical images show dark

*features on the sample that aided with rotating and repositioning the sample. 3e) and f) show RGB composite images of the z', y' and x' piezoresponses estimated with the sample at 0 degrees (3e)) and 90 degrees(3f)) respectively. As with Figure 2 above, the images were rotated after processing to all appear in the original (Figure 3a)) frame of reference to make comparison easier. These data were acquired with a Vero QPDI interferometric system. All scans were 4µm.*

## 4. Discussion

The workflow illustrated above provides a built-in check on tip wear, sample changes and other cheges during the measuremnt process. Specifically, the four point measurement approach decribed in step #1 above results in producing two independent estimates of the vertical (z') piezoresponse in addition to the longitudinal (x') and lateral (y') measurements. This provides for a built-in check on at least the vertical component. Then, can repeating the same workflow at different sample orientations (specifically 0 and 90 degrees) provides another set of estimates, allowing a test of the measurement confidence. Figure S4 in the supplemental materials shows an example where there were changes in the sample during the measurement workflow, yielding lower confidence estimates.

For isotropic solids, we expect that Young's modulus $E$ is related to the shear modulus $G$ though the relationship $E = G(1 + 2\mu)$, where $\mu$ is the Poisson's ratio. Furthermore, for a probe in Hertzian contact with the surface, the normal stiffness $k_z$ is given by $k_z = 2aE$, where a is the contact radius. Similarly, the lateral stiffness is given by $k_{lat} = 8aG$.[49] The ratio of these two stiffnesses is given by $k_{lat}/k_z = 4/(1 + 2\mu)$. Since the Poisson ratio of both PZT and BFO is $\mu \approx 0.3$, then $k_{lat}/k_z \approx 2.5$ for both materials. If we assume the tip displacement is inversely proportional to the tip-sample stiffness, then $d_{eff,z}/d_{eff,lat} \approx \pm 2.5 \pm 0.5$, in reasonable agreement with our experimental observations. See supplemental Figure S3 for plots of the ratios for both Figures 2 and 3. The error estimate results from applying the error budget summarized in Table S1, supported by Figures S7-S9. One source of error that is not considered in this paper is lateral and/or longitudinal flexing of the conical tip of the AFM probe itself.[25] This effect has could undermine the above conclusions significantly, although preliminary measurements indicate it is not significant for the probes we used in this work.[50]

Quantitative electromechanical response measurements, a long-standing goal for the PFM community, require consideration of the electro-elastic fields generated inside a material due to the contact of a sharp, conductive tip on a piezoelectric sample. Two decades ago, Kalinin at al.[51], [52] discussed the so-called strong indentation case, where the coupled electro-elastic problem for piezoelectric indentation is solved to obtain the electric field and strain distribution in the ferroelectric material. Specifically, they predicted a $d_{eff,z'} \approx 12\text{pm/V}$ for the hard indentation limit sensitivity of lithium niobate. While these analytical solutions have contributed to our understanding of PFM contact mechanics and imaging mechanisms, experimental verification of the theoretical predictions have remained elusive until recently, where, through a series of careful interferometric measurements, we were able to experimentally show that in the limit of high indentation forces, $d_{eff,z'} \to 12\text{pm/V}$ for stoichiometric lithium niobate.[28] One of the goals for this work was to test the self-consistency of vector measurements as the sample was rotated. Generally, samples show agreement between the vertical components measured at different in-plane orientation angles (Equations (5) and (6)) as long as they are time-stable, as was the case for the data shown in Figures 2 and 3. These results validate our cantilever model with a linear beam near the end (see Figure 1).

If the measurements are more invasive, if there is tip-wear or other changes during the workflow described here, it will lead to increasing uncertainty in the estimated components. An example of this is shown in supplemental Figure S4 from our initial vector mapping attempts. In this case, we used a relatively large *ac excitation voltage,* $V_{ac} = 5V$ and a load of $\approx 1\mu N$. That large values may have caused some domain mobility during the process of acquiring the images, resulting in very highly variable estimates for the in-plane components ($d_{eff,\ x'}$ and $d_{eff,\ y'}$) in particular, implying there is a dependence on load and friction that may warrant future study. To mitigate this behavior, the data of Figure 3 were acquired at both a lower excitation value $V_{ac} = 1V$ and a lower load of $\approx 200nN$, resulting in much more consistent results. We are hopeful that the quantitative approach we have presented here will enable future progress in understanding the rich interplay of bias, load, tip-sample physics, environmental factors and intrinsic material properties on electromechanical responses.

We considered several sources of error associated with positioning of the spot on the cantilever and with the estimation of the cantilever tip height. These estimates are described more fully in the supplemental materials, specifically Figures S7-S9 and in Table S1. Briefly, the uncertainties for the three components are $\Delta d_{eff,\ x'}/d_{nom,\ x'} \approx 13\%$, $\Delta d_{eff,\ y'}/d_{nom,\ y'} \approx 13\%$, and $\Delta d_{eff,\ z'}/d_{nom,\ z'} \approx 6\%$. The in-plane estimates are most strongly dependent on the lateral separation of the spots (~3%) and tip height (~10%). The vertical uncertainties are dominated by errors in the spot separation (~3%) or tip location along the long axis of the cantilever (3%). For the error budget, we conservatively assumed all dimensional uncertainties were 2um. This conservative value is much larger than the 200nm resolution of the encoders in the spot positioning stage and is also consistent with independent comparisons of the optical approach to estimating the tip height described here with CD-SEM measurements.[50]

## 5. Conclusions

The sequential interferometric piezoresponse force microscopy (PFM) method introduced here offers a significant advancement in the ability to quantitatively measure the full three-dimensional electromechanical response of piezoelectric materials without the need for sample rotation. By leveraging the reduced noise floor of our interferometric displacement sensors, this approach enables precise and reproducible in-plane and vertical strain measurements. The findings, including a systematic difference in the sensitivities of in-plane and vertical responses, agree well with theoretical models, validating the robustness of this method. Additionally, the ability to map full 3D responses in a time-efficient manner, while minimizing labor and errors associated with sample repositioning, marks a considerable improvement over traditional angle-resolved PFM techniques. This work lays the foundation for more accurate and comprehensive three-dimensional characterization of piezoelectric and ferroelectric materials. Future studies will focus on exploring the complex tip-sample interactions further, particularly with regard to load and frictional effects, to enhance the precision of these measurements even more. Furthermore, we expect that this general formalism will be extended to other AFM measurement modes including a wide variety of nanomechanical measurements.

**ASSOCIATED CONTENT**

**Supporting Information**

**AUTHOR INFORMATION**

**Corresponding Author**

Roger.proksch@oxinst.com

**Notes**

The authors declare no competing financial interests.

**ACKNOWLEDGMENTS**

RP thanks Gustau Catalan, Sergei Kalinin and Aleks Labuda for useful and informative discussions. Ramamoorthy Ramesh supplied the BFO sample.

# REFERENCES

1. J. Liu, Z. N. Bao, Y. Cui, E. J. Dufek, J. B. Goodenough, P. Khalifah, Q. Y. Li, B. Y. Liaw, P. Liu, A. Manthiram, Y. S. Meng, V. R. Subramanian, M. F. Toney, V. V. Viswanathan, M. S. Whittingham, J. Xiao, W. Xu, J. H. Yang, X. Q. Yang and J. G. Zhang, Nature Energy **4** (3), 180-186 (2019).
2. J. B. Pang, R. G. Mendes, A. Bachmatiuk, L. Zhao, H. Q. Ta, T. Gemming, H. Liu, Z. F. Liu and M. H. Rummeli, Chemical Society Reviews **48** (1), 72-133 (2019).
3. K. He, C. Poole, K. F. Mak and J. Shan, Nano Letters **13** (6), 2931-2936 (2013).
4. D. Akinwande, C. J. Brennan, J. S. Bunch, P. Egberts, J. R. Felts, H. J. Gao, R. Huang, J. S. Kim, T. Li, Y. Li, K. M. Liechti, N. S. Lu, H. S. Park, E. J. Reed, P. Wang, B. I. Yakobson, T. Zhang, Y. W. Zhang, Y. Zhou and Y. Zhu, Extreme Mechanics Letters **13**, 42-77 (2017).
5. B. C. Wyatt, A. Rosenkranz and B. Anasori, Advanced Materials **33** (17) (2021).
6. S. Slesazeck and T. Mikolajick, Nanotechnology **30** (35) (2019).
7. C. Wu, T. W. Kim, H. Y. Choi, D. B. Strukov and J. J. Yang, Nature Communications **8** (2017).
8. R. Ramesh, S. Salahuddin, S. Datta, C. H. Diaz, D. E. Nikonov, I. A. Young, D. Ham, M. F. Chang, W. S. Khwa, A. S. Lele, C. Binek, Y. L. Huang, Y. C. Sun, Y. H. Chu, B. Prasad, M. Hoffmann, J. M. Hu, Z. Yao, L. Bellaiche, P. Wu, J. Cai, J. Appenzeller, S. Datta, K. Y. Camsari, J. Kwon, J. A. C. Incorvia, I. Asselberghs, F. Ciubotaru, S. Couet, C. Adelmann, Y. Zheng, A. M. Lindenberg, P. G. Evans, P. Ercius and I. P. Radu, Apl Materials **12** (9) (2024).
9. F. Guinea, M. I. Katsnelson and A. K. Geim, Nature Physics **6** (1), 30-33 (2010).
10. J. M. Kim, M. F. Haque, E. Y. Hsieh, S. M. Nahid, I. Zarin, K. Y. Jeong, J. P. So, H. G. Park and S. Nam, Advanced Materials **35** (27) (2023).
11. G. Binnig, C. F. Quate and C. Gerber, Physical review letters **56** (9), 930 (1986).
12. G. Meyer and N. M. Amer, Applied Physics Letters **57** (20), 2089-2091 (1990).
13. G. Meyer and N. M. Amer, Applied Physics Letters **53** (12), 1045-1047 (1988).
14. M. Munz, Journal of Physics D-Applied Physics **43** (6) (2010).
15. K. L. Kim and J. E. Huber, Applied Physics Letters **104** (12) (2014).
16. K. Chu and C. H. Yang, Review of Scientific Instruments **89** (12) (2018).
17. B. Kim, F. P. Barrows, Y. Sharma, R. S. Katiyar, C. Phatak, A. K. Petford-Long, S. Jeon and S. Hong, Scientific Reports **8** (2018).
18. K. L. Kim and J. E. Huber, Review of Scientific Instruments **86** (1) (2015).
19. M. Park, S. Hong, J. Kim, J. Hong and K. No, Applied Physics Letters **99** (14) (2011).
20. M. Park, S. Hong, J. A. Klug, M. J. Bedzyk, O. Auciello, K. No and A. Petford-Long, Applied Physics Letters **97** (11) (2010).
21. D. Seol, H. Taniguchi, J. Y. Hwang, M. Itoh, H. Shin, S. W. Kim and Y. Kim, Nanoscale **7** (27), 11561-11565 (2015).
22. F. P. Zhuo and C. H. Yang, Physical Review B **102** (21) (2020).
23. N. Alyabyeva, A. Ouvrard, I. Lindfors-Vrejoiu, A. Kolomiytsev, M. Solodovnik, O. Ageev and D. McGrouther, Physical Review Materials **2** (6) (2018).
24. L. He, J. W. Meng, B. Y. Zhao, J. Jiang, W. P. Geng and A. Q. Jiang, Ferroelectrics **492** (1), 59-68 (2016).
25. D. O. Alikin, A. S. Abramov, M. S. Kosobokov, L. V. Gimadeeva, K. N. Romanyuk, V. Slabov, V. Y. Shur and A. L. Kholkin, Ferroelectrics **559** (1), 15-21 (2020).
26. D. O. Alikin, L. V. Gimadeeva, A. V. Ankudinov, Q. Hu, V. Y. Shur and A. L. Kholkin, Applied Surface Science **543** (2021).
27. A. Labuda and R. Proksch, Applied Physics Letters **106** (25), 253103 (2015).
28. R. Proksch, R. Wagner and J. Lefever, Journal of Applied Physics **135** (3) (2024).
29. Y. Martin, C. C. Williams and H. K. Wickramasinghe, Journal of Applied Physics **61** (10), 4723-4729 (1987).
30. D. Rugar, H. J. Mamin, R. Erlandsson, J. E. Stern and B. D. Terris, Review of Scientific Instruments **59** (11), 2337-2340 (1988).
31. B. W. Hoogenboom, P. Frederix, D. Fotiadis, H. J. Hug and A. Engel, Nanotechnology **19** (38) (2008).
32. Y. T. Liu, L. Collins, R. Proksch, S. Kim, B. R. Watson, B. Doughty, T. R. Calhoun, M. Ahmadi, A. V. Ievlev, S. Jesse, S. T. Retterer, A. Belianinov, K. Xiao, J. S. Huang, B. G. Sumpter, S. V. Kalinin, B. Hu and O. S. Ovchinnikova, Nature Materials **17** (11), 1013-+ (2018).
33. S. S. Cheema, D. Kwon, N. Shanker, R. dos Reis, S. L. Hsu, J. Xiao, H. G. Zhang, R. Wagner, A. Datar, M. R. McCarter, C. R. Serrao, A. K. Yadav, G. Karbasian, C. H. Hsu, A. J. Tan, L. C. Wang, V. Thakare, X. Zhang, A. Mehta, E. Karapetrova, R. V. Chopdekar, P. Shafer, E. Arenholz, C. M. Hu, R. Proksch, R. Ramesh, J. Ciston and S. Salahuddin, Nature **580** (7804), 478-+ (2020).
34. L. Collins, Y. T. Liu, O. S. Ovchinnikova and R. Proksch, Acs Nano **13** (7), 8055-8066 (2019).
35. Z. M. Zhang, S. L. Hsu, V. A. Stoica, H. Paik, E. Parsonnet, A. Qualls, J. J. Wang, L. Xie, M. Kumari, S. Das, Z. N. Leng, M. McBriarty, R. Proksch, A. Gruverman, D. G. Schlom, L. Q. Chen, S. Salahuddin, L. W. Martin and R. Ramesh, Advanced Materials **33** (10) (2021).
36. A. Lipatov, P. Chaudhary, Z. Guan, H. D. Lu, G. Li, O. Crégut, K. D. Dorkenoo, R. Proksch, S. Cherifi-Hertel, D. F. Shao, E. Y. Tsymbal, J. Iñiguez, A. Sinitskii and A. Gruverman, Npj 2d Materials and Applications **6** (1) (2022).
37. H. D. Lu, H. Aramberri, A. Lipatov, R. Proksch, A. Sinitskii, J. Iñiguez and A. Gruverman, Acs Materials Letters **5** (11), 3136-3141 (2023).
38. https://afm.oxinst.com/products/vero-family-afms/vero-afm, (2024).
39. M. Abplanalp, L. M. Eng and P. Gunter, Applied Physics a-Materials Science & Processing **66**, S231-S234 (1998).
40. S. V. Kalinin, B. J. Rodriguez, S. Jesse, J. Shin, A. P. Baddorf, P. Gupta, H. Jain, D. B. Williams and A. Gruverman, Microscopy and Microanalysis **12** (3), 206-220 (2006).
41. R. Nath, S. Hong, J. A. Klug, A. Imre, M. J. Bedzyk, R. S. Katiyar and O. Auciello, Applied Physics Letters **96** (16) (2010).
42. T. Jungk, A. Hoffmann and E. Soergel, Journal of Microscopy-Oxford **227** (1), 72-78 (2007).

43. N. Balke, S. Jesse, P. Yu, B. Carmichael, S. V. Kalinin and A. Tselev, Nanotechnology **27** (42) (2016).
44. S. M. Neumayer, S. Saremi, L. W. Martin, L. Collins, A. Tselev, S. Jesse, S. V. Kalinin and N. Balke, Journal of Applied Physics **128** (17) (2020).
45. B. D. Huey, C. Ramanujan, M. Bobji, J. Blendell, G. White, R. Szoszkiewicz and A. Kulik, Journal of Electroceramics **13** (1-3), 287-291 (2004).
46. A. Hoffmann, T. Jungk and E. Soergel, Review of Scientific Instruments **78** (1) (2007).
47. F. Zavaliche, R. R. Das, D. M. Kim, C. B. Eom, S. Y. Yang, P. Shafer and R. Ramesh, Applied Physics Letters **87** (18) (2005).
48. M. Park, S. Hong, J. A. Klug, M. J. Bedzyk, O. Auciello, K. No and A. Petford-Long, Applied Physics Letters **97** (11) (2010).
49. A. Makagon, M. Kachanov, S. V. Kalinin and E. Karapetian, Physical Review B **76** (6) (2007).
50. J. Lefever, A. Labuda and R. Proksch, Submitted (2025).
51. S. V. Kalinin, E. Karapetian and M. Kachanov, Physical Review B **70** (18) (2004).
52. S. V. Kalinin, E. A. Eliseev and A. N. Morozovska, Applied Physics Letters **88** (23) (2006).